\font\tenbg=cmmib10 at 10pt
\def \rvecmu{{\hbox{\tenbg\char'026}}}

\documentclass[useAMS]{mn2e}
\usepackage[dvipdfm]{graphicx}

\title{Spinning-Down of Moving Magnetars in the Propeller Regime}

\author[O.D. Toropina et al.]
{O.D. Toropina,$^1$\thanks{E-mail:toropina@iki.rssi.ru}, M.M.
Romanova,$^2$, and R.V.E. Lovelace,$^{2,3}$\\
$^1$ Space Research Institute, Russian Academy of Sciences,
Profsoyuznaya 84/32, Moscow 117997, Russia\\
$^2$ Department of Astronomy, Cornell University, Ithaca, NY
14853-6801\\
$^3$
Department of Applied and Engineering Physics, Cornell University,
Ithaca, NY 14853-6801}

\date{\today}
\pagerange{\pageref{firstpage}--\pageref{lastpage}}
\pubyear{2006}

\begin{document}

\label{firstpage}

\maketitle

\begin{abstract}

    We use axisymmetric magnetohydrodynamic
simulations  to investigate  the  spinning-down of magnetars
rotating in the propeller regime and moving supersonically through
the interstellar medium.
   The simulations indicate that magnetars spin-down rapidly due to
this interaction,  faster than for the case of a non-moving star.
 From many simulation runs we have derived an
approximate scaling laws for the angular momentum loss rate,
$\dot{L} \propto-~\eta_m^{0.3}\mu^{0.6}\rho^{0.8}{\cal M}^{-0.4}
\Omega_*^{1.5}$, where $\rho$ is the  density of the interstellar
medium, ${\cal M}$ is Mach number, $\mu$ is the star's magnetic
moment, $\Omega_*$ is its angular velocity, and $\eta_m$ is magnetic
diffusivity.
   A magnetar with a surface magnetic field of $10^{13} - 10^{15}$ G
is found to spin-down to a period $P > 10^5-10^6$ s in  $\sim 10^4 -
10^5$ years. There is however  uncertainty about the value of the
magnetic diffusivity so that the time-scale may be longer.
 We discuss this
model in respect of Soft Gamma Repeaters (SGRs) and the isolated
neutron star candidate RXJ1856.5-3754.

\begin{keywords}  neutron stars --- magnetars --- magnetic field
--- MHD
\end{keywords}

\end{abstract}

\section{Introduction}

Some neutron stars referred to as
``magnetars'' have unusually large
magnetic fields, $B\sim 10^{13}-10^{15}
~{\rm G}$ (Duncan \& Thompson
1992; Thompson \& Duncan 1995).
  Possible candidates for magnetars
include anomalous X-ray pulsars
and soft gamma-ray repeaters (SGRs)
(Kulkarni \& Frail 1993; Kouveliotou et al. 1994;
Hurley et al.
1999).
     These objects are associated
with  supernovae remnants and
hence are relatively young
(Vasist \& Gotthelf 1997; Kouveliotou et
al. 1998).
  Only a few candidates for magnetars
have been found so far.
  The estimated birthrate of magnetars
is $\sim  10\% $ of ordinary pulsars (Kulkarni \& Frail 1993;
Kouveliotou et al. 1994, 1999) so that there might be many more
magnetars which are presently invisible.
  Their ``visibility" depends
on a number of factors.
One important factor is the rate
of the star's spin-down.
   If magnetars
spin-down rapidly to very long
periods, then one will not detect spin
modulated variability during
flares.
   Consequently the identification
of the variability with a rotating neutron star would be
more difficult.

\begin{figure*}
\includegraphics[scale=.8]{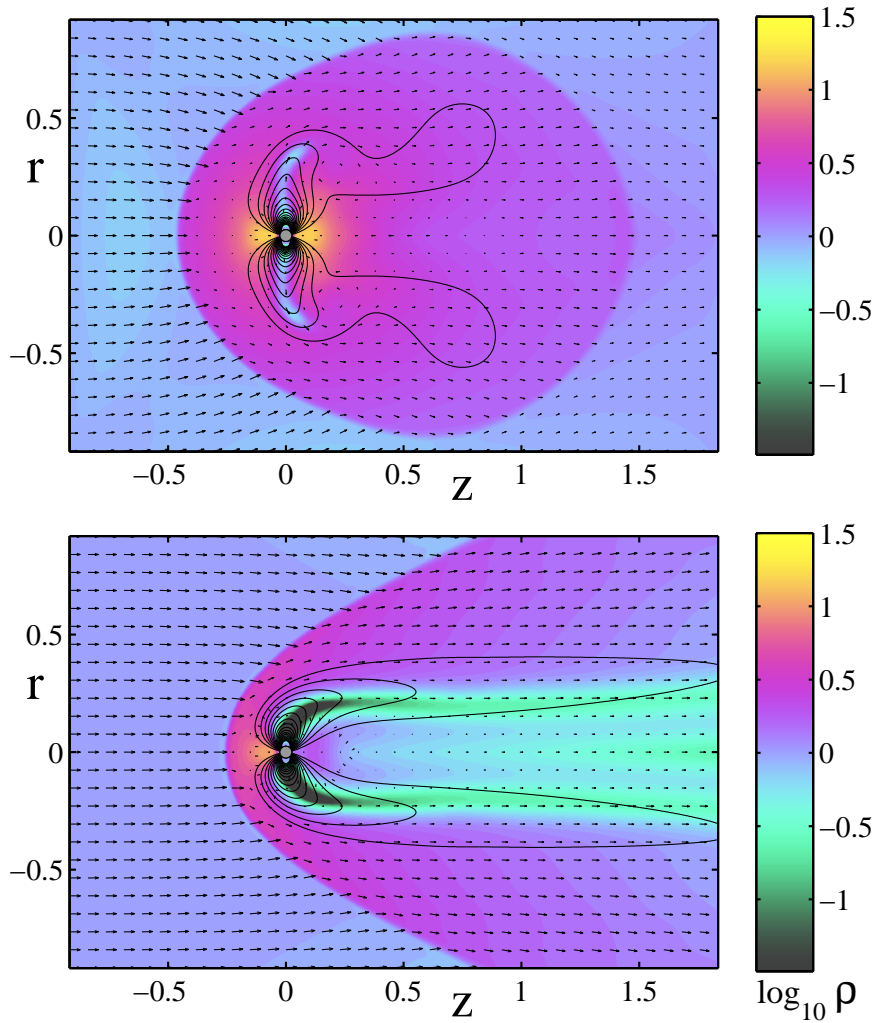}
\caption{Matter flow around a strongly
magnetized star rotating in the
propeller regime and propagating
through the interstellar medium
with Mach numbers ${\cal M}=1$ (top
panel) and
  ${\cal M}=3$ (bottom panel).
    Other parameters correspond to the main case.
The background represents the
logarithm of density and the length
  of the arrows is proportional
to the poloidal velocity.
    The solid lines are magnetic field lines.
The dashed solid line shows the Alfv\'en surface.
    Distances are
measured in units of the Bondi radius.
   Time corresponds to $50$ rotation
periods of of the star.}
\label{Figure 1}
\end{figure*}

During the pulsar stage of evolution,
magnetars spin down much more
rapidly than ordinary pulsars.
 Consequently, they pass through
their pulsar stage much faster,
in $\sim 10^4$ years (Thompson \&
Duncan 1995).
  When the light cylinder radius becomes larger than
magnetospheric radius $r_m$,
the relativistic wind is suppressed by
the inflowing matter (Shvartsman 1970)
and the star enters the
propeller regime where the spin-down
is due to the interaction of
the star's rotating field with the
interstellar medium (ISM) (Davidson \&
Ostriker 1973; Illarionov \& Sunyaev 1975).
  Magnetars with
velocities $v > 100-200$ km/s  interact directly with
the ISM. That is, the magnetospheric radius is larger than
gravitational capture radius (Harding \& Leventhal 1992; Rutledge
2001; Toropina et al. 2001).
    In the propeller regime the rapidly rotating
magnetosphere interacts strongly with the supersonically
inflowing ISM.

The spin-down rate of supersonically
moving  magnetars in the
propeller regime has been estimated earlier,
but different authors
have obtained rather different results.
    For example,
Rutledge (2001) estimates a
spin-down time of $\sim 4 \times
10^9$ yr for a neutron star with a surface
magnetic field $B = 10^{15}~{\rm G}$ and
velocity $v = 300~{\rm km/s}$.
   On the other hand, Mori and Ruderman
(2003) estimate that a magnetar
spins-down to periods greater
than $10^4$ s within $\sim 5 \times 10^5$ years.
   Mori and Ruderman put forward this
model as an explanation of the isolated neutron star (INS) candidate
RX J1856.5-3754 which does not show variability.

The propeller  stage of evolution has been investigated in
non-magnetar cases both theoretically (e.g., Illarionov \& Sunyaev
1975; Davies, Fabian \& Pringle 1979; Davies \& Pringle 1981;
Lovelace, Romanova \& Bisnovatyi-Kogan 1999; Ikhsanov 2002;
Rappaport, Fregeau, \& Spruit, 2004) and with MHD simulations (Wang
\& Robertson 1985; Romanova et al. 2003; Romanova et al. 2004;
Romanova et al. 2005; Ustyugova et al. 2006).
  However, only limited theoretical work has been
done for magnetar-type propellers,
which propagate through the ISM
supersonically (Rutledge 2001; Mori \& Ruderman 2003).
 No MHD simulations of this flow regime
have been done previously.

    This work presents the first
numerical  simulations of the  interaction of supersonic, fast
rotating magnetars with the ISM.
   Earlier, we investigated supersonic
propagation of {\it non-rotating} strongly
  magnetized stars through the ISM (Toropina et al.
2001). The present simulations
are analogous to those in Toropina et al.
(2001). However, here the star rotates in the propeller regime.
 The main objective of this work
is to determine the dependences of spin-down rate of the star
on the different variables
 and to estimate corresponding time-scale of the spin-down.

Sections 2 and 3  describe the physical situation
and simulation model.
  In \S 4, we discuss the results  of our  simulations.
   In \S 5, we apply our results to magnetars and  in \S 6
we discuss
   possible magnetar candidates.
Conclusions are given in Section 7.

\begin{figure*}
\includegraphics[scale=.7]{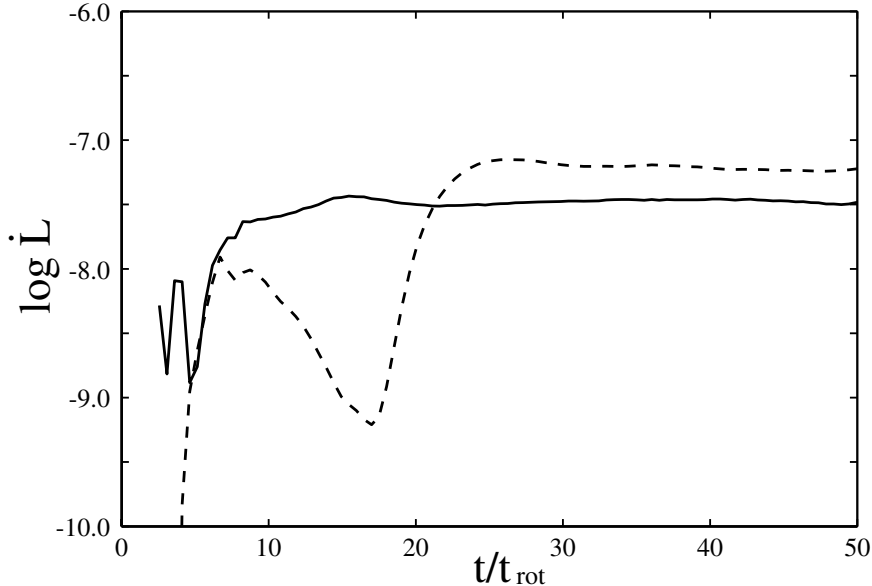}
\caption{Temporal variation of the
total angular momentum flux from the star
obtained by integrating over a surface
surrounding the star
(solid line) and the flux through  a surface
across the magnetotail at $z=0.6$ (dotted line).}
\label{Figure 2}
\end{figure*}

\begin{figure*}
\includegraphics[scale=.8]{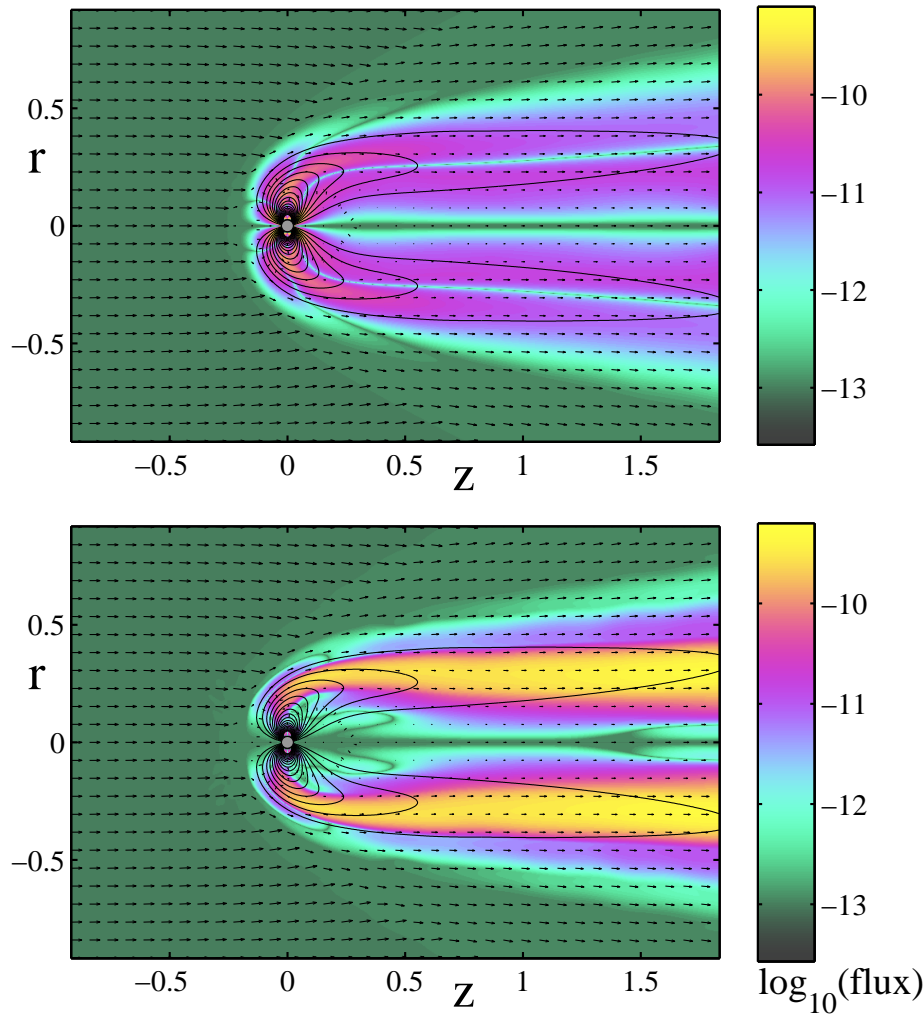}
\caption{Distribution of the angular
momentum fluxes in the magnetotail for our  main case.
    The color
background shows the specific angular
momentum carried by the magnetic field
(top panel) and by the matter (bottom panel).
The solid lines are magnetic
field lines.}
\label{Figure 3}
\end{figure*}

\begin{figure*}
\includegraphics[scale=.8]{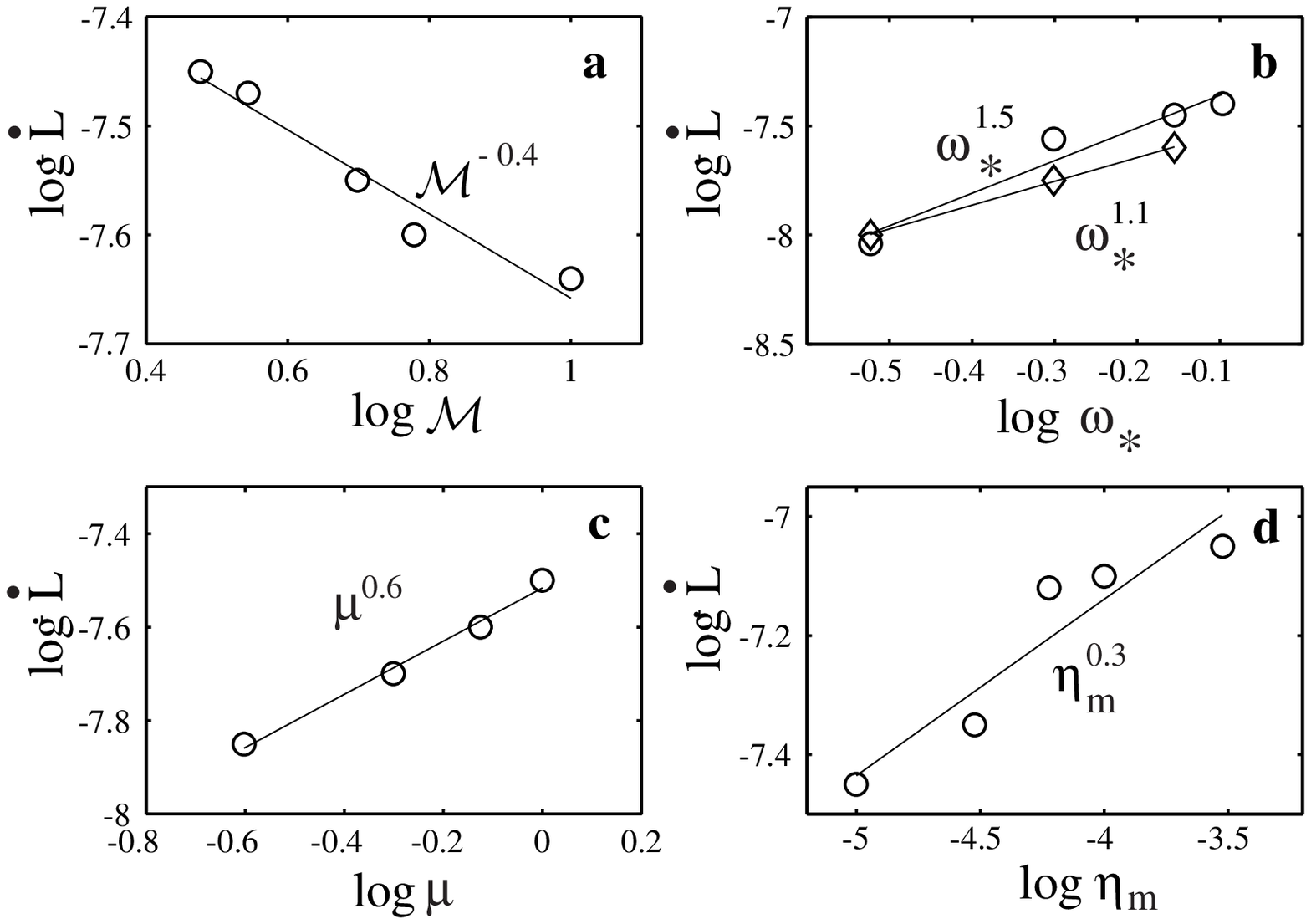}
\caption{Dependence of the angular
momentum flux on different parameters,
(a) the Mach number ${\cal
M}$, (b) the angular velocity of
the star $\omega_*\equiv \Omega_*/\Omega_K$,
(c) the magnetic moment $\mu$, and (d) the magnetic
diffusivity
$\tilde{\eta}_m$.} \label{Figure 4}
\end{figure*}

\section{Physics of the ``Propeller" Regime of a Moving Magnetar}

      Here, we consider the spin-down of
a highly magnetized rotating neutron star or ``magnetar'' moving at
a velocity $v$ faster than the local sound speed of the interstellar
medium;  that is, $v > c_s$, where $c_s \approx 10^6{~\rm cm/s}$ for
a $10^4~{\rm K}$ hydrogen plasma.
    Theory and simulations of the spin-down for
the case of Bondi accretion to a non-moving, rotating magnetized
star in the propeller regime was discussed earlier by Romanova et
al. (2003).
       A general
discussion and review of theory of the spin-down of moving rotating
magnetars  is given by Mori and Ruderman (2003).

     We first discuss the different length
scales in this problem.   We let $r_m$ denote the characteristic
size of the star's magnetosphere with the region inside $r_m$
magnetically dominated and corotating with the star at angular rate
$\Omega_*$.

     The standoff distance of the bow shock
formed by the interaction of the moving ISM with the star's magnetic
field is at a distance of the order of
$$
r_{\rm st} \equiv \left({\mu^2 \over 4\pi \rho v^2} \right)^{1/6}
\approx 2.8\times 10^{12} \left({\mu_{33}^{1/3} \over n^{1/6}
v_{7}^{1/3}}\right){~\rm cm}~, \eqno(1)
$$
from the center of the star. Here, $\mu_{33} = \mu/10^{33}$Gcm$^3$
is the star's magnetic moment, $n$ is the number density of the ISM
in H/cm$^3$, and $v_{7} \equiv v/(10^7$cm/s).
    Note that the  magnetic field at
the star's surface is $B = \mu/ R_*^3 =10^{15}\mu_{33}(10{\rm
km}/R_*)^3$ G, where $R_*$ is the star's radius.  We assume that
$r_{st}  < r_m$.

      In the propeller regime the magnetospheric
radius $r_m$ is larger than the corotation radius
$$
r_{\rm cr}\equiv (GM/\Omega_*^2)^{1/3}\approx 1.7\times 10^{10}
P_{3}^{2/3}~{\rm cm}~, \eqno(2)
$$
where $P$ is the rotation period of the star and $P_3 \equiv
P/10^3$ s. We assume  $M=1.4M_\odot$ here and subsequently.
Consequently, the incoming matter is flung away from the star in
the equatorial plane.

A further characteristic length is
$$
   r_v \equiv v/\Omega_* =1.5\times 10^9 v_7 P_3~{\rm cm}~,
\eqno(3)
$$

    This is the distance the star moves in
a rotation period divided by $2\pi$.
    A related length is
$$
r_s \equiv c_s/\Omega_*~, \eqno(4)
$$
where $c_s$ is the sound speed in ambient ISM.   Other
characteristic lengths can be constructed from combinations of
equations (1) - (4).  For example, the Bondi radius is given by
$r_{\rm B} = r_{\rm cr}^3/r_s^2$, and the Bondi-Hoyle radius by
$r_{\rm BH} = r_{\rm cr}^3/r_v^2$.

       Additionally,
there is a length scale associated with
the magnetic diffusivity $\eta$
$$
r_\eta \equiv (\eta/\Omega_*)^{1/2}~.
 \eqno(5)
$$
Roughly, $r_\eta$ is the distance the magnetic field can diffuse in
a rotation period of the star.
 These lengths are expected to be small compared
with the lengths (1) - (4). Note that magnetic diffusivity is much
more important in this problem, compared to viscosity, so that we do
not take into account viscosity in the analysis below.

     Note also that we assume the light cylinder radius,
$$
r_{\rm L} = {cP / 2\pi} \approx 4.8\times 10^{12} P_3~{\rm cm}~,
\eqno(6)
$$
is larger than the standoff distance $r_{\rm st}$.

\subsection{Torque Derived from Dimensional Analysis}

     The rate of spin-down  of the star is
estimated as $dL/dt=Id\Omega_*/dt \approx - 4\pi r_m^2
fr_mB_m^2/4\pi$, which is the magnetic torque at the radius $r_m$,
where $I$ is the star's moment of inertia, and $f$ is the fraction
of the solid angle where this torque acts.
   We have assumed that $|B_\phi| \approx B_m$
at $r_m$.
    Thus
$$
{d L \over dt}=I { d\Omega_* \over dt} \approx - {f\mu^2  \over
r_{m}^3}~. \eqno(7)
$$
We take an empirical approach to determining the dependences of the
magnetospheric radius $r_m$ on the different physical parameters
$$
r_m=r_m(\rho,~\mu, ~v,~c_s, ~\Omega_*,~ \eta)~, \eqno(8)
$$
as suggested by Mori and Ruderman (2003).
     We write
$$
r_m =(r_{\rm st})^\alpha (r_{\rm cr})^\beta (r_\eta)^\gamma
 (r_s)^\epsilon (r_v)^{s}~, \eqno(9)
$$
where $s\equiv 1-\alpha-\beta -\gamma-\epsilon$. Consequently,
$$
{dL\over dt}\propto ~-~\rho^{\alpha/2}
\mu^{2-\alpha}\eta^{-3\gamma/2} c_s^{-3\epsilon} v^{3s-\alpha}
\Omega_*^{q}~, \eqno(10)
$$
where $q=2\beta+3\gamma/2+3\epsilon+3s$.
    In \S 4  we discuss the values of the
exponents suggested by our simulations.

\subsection{Torque Derived from  Analytical Models}

  There is a wide variety of theories
which predict different
dependencies for the spin-down torque.
  The dependencies are summarized in
tables by Davies, et al. (1979)
and by MR03.
  For example, from the table in MR03 we see
that the power law dependence on
$\Omega$ varies strongly, from
$n_\Omega=-1$ up to 2; the power law
dependence on $\mu$ varies less
dramatically from $n_\mu=2/3$ up to $5/3$.
   The dependence on  the star's
velocity  varies dramatically from
$n_v=-5/3$ to $7/3$.
   The dependence on the
density of the ISM  varies from $n_\rho=1/6$
to $2/3$.
  The torque obtained from the
different models differs by
several orders of magnitude.
   The spin-down time scales vary correspondingly.
  One of the reasons for the differences
is that in different situations (disk accretion, wind
accretion to the non-moving or moving pulsar) the physics of
interaction is different.
    Only a few papers deal with the situation
considered in this paper where the magnetospheric
radius is larger than
the gravitational capture radius.
   One  consequences of this
is that the magnetospheric radius $r_m$ is of the order
of the stand-off radius $r_{st}$.
  Below, we propose a simple physical theory
which is closest to  our particular problem.

\subsection{Torque Derived from Simple Physical Model}

Here, we give a simple model for the spin
down of a moving propeller star.
We assume that the
magnetospheric radius is larger than the accretion radius so
that the magnetosphere is an obstacle for the interstellar
matter.
 Matter stops at the stand-off distance $r_{st}$ determined from the
balance of the ram pressure of the ISM
and the magnetic pressure of the star and is given by
equation (1).
  The mass flow crossing the area $\pi r_{st}^2$ is
$$
\dot M =\pi r_{st}^2\rho v  ~.
$$
We assume that a fraction $f=k[\eta/(r_{st} v)]^s$ of this mass flux
participate in spinning-down of the star, where $Re_m = r_{st}
v/\eta$ is a magnetic Reynolds number and $k$ and $s$ are constants.
   The factor $f$ is determined by the penetration of
the interstellar plasma into the star's magnetosphere.
   The larger the penetration the larger the transfer
of angular momentum from the star to the ISM. Noting that $\Omega_*
r_{st}^2$ is the specific angular momentum of the matter at a
distance $r_{st}$ we obtain
$$
{dL \over dt} = -k \left(\eta \over r_{st} v\right)^s \dot M
\Omega_* r_{st}^2\approx  - k \eta^s \rho^{(s+2)\over 6}
\mu^{(4-s)\over 3} v^{-(1+2s)\over 3} \Omega_*.
$$
As an illustration  consider $s=0.3$ which is suggested by
our simulations (see Figure 4d).
   We then find
$$
   {dL \over dt} = - k \eta^{0.3}\rho^{0.4}
\mu^{1.2} \Omega_*v^{-0.5}~ .
\eqno(11)
$$
This model
indicates  that the torque  increases with $\mu$,  $\rho$ and
   $\Omega_*$, but it
but  decreases with increasing velocity $v$.

\subsection{Transition from Pulsar to Propeller Stages}

Equating the standoff  distance $r_{st}$ to
the light cylinder radius $r_L$ gives an estimate
of the transition from the pulsar  stage to the propeller
stage,
$$
P_{\rm trans} = 5.6\times 10^2  \left({\mu_{33}^{1/3} \over n^{1/6}
v_{7}^{1/3}}\right){~\rm s}~.
\eqno(12)
$$
The propeller stage will end when the corotation
radius $r_{cr}$ is larger than the magnetospheric radius
$r_{st}$.
  This occurs for
$$
P = 2.1\times 10^6 \left({\mu_{33}^{1/2} \over n^{1/4}
v_{7}^{1/2}}\right){~\rm s}~.
\eqno(13)
$$
Thus, a magnetar with  field  $B=10^{15}~{\rm G}$ propagating
through the ISM supersonically will be in the propeller regime for
rotation periods $560 s < P < 2\times 10^6 s$. Our model applies
only for this range.


\section{Numerical Model}

We  investigate the interaction of fast moving rotating magnetized
star with the ISM using an axisymmetric, resistive MHD code. The
code incorporates the methods of local iterations and
flux-corrected-transport (Zhukov, Zabrodin, \& Feodoritova 1993).
The flow is described by the resistive MHD equations:
$$
    {\partial \rho \over
    \partial t}+
    {\bf \nabla}{\bf \cdot}
\left(\rho~{\bf v}\right) =0{~ ,}
$$
$$
\rho\frac{\partial{\bf v}} {\partial t}
              +\rho ({\bf v}
\cdot{\bf \nabla}){\bf v}=
    -{\bf \nabla}p+
{1 \over c}{\bf J}\times {\bf B} + {\bf F}^{g}{ ~,}
$$
$$
    \frac{\partial {\bf B}}
{\partial t}
=
    {\bf \nabla}{\bf \times}
\left({\bf v}{\bf \times} {\bf B}\right)
    +
    \frac{c^2}{4\pi\sigma}
\nabla^2{\bf B} {~,}
$$
$$
    \frac{\partial (\rho\varepsilon)
}{\partial t}+
    {\bf \nabla}\cdot \left(\rho
\varepsilon{\bf v}\right)
=
    -p\nabla{\bf \cdot}
{\bf v} +\frac{{\bf J}^2} {\sigma}{~.} \eqno(14)
$$
We assume axisymmetry $(\partial/\partial \phi =0)$,
but calculate
all three components of velocity
and magnetic field ${\bf v}$ and
${\bf B}$.
    The equation of state is taken
to be that for an ideal gas,
$p=(\gamma-1)\rho \varepsilon$,
with specific heat ratio
$\gamma=5/3$.
      The equations incorporate Ohm's law ${\bf
J}=\sigma({\bf E}+{\bf v}  \times {\bf B}/c)$,
where $\sigma$ is
the electrical conductivity.
     The associated magnetic diffusivity,
$\eta_m \equiv c^2\!/(4\pi\sigma)$,
is considered to be a constant
within the computational region. In
equation (2) the gravitational force,
${\bf F}^{g} = -GM\!\rho{\bf
R}/\!R^3$, is due to the central star.

We use a cylindrical, inertial
coordinate system $\left(r,\phi,z\right)$
with the $z-$ axis parallel to the
star's dipole moment ${\rvecmu}$ and
rotation axis ${\bf \Omega}$.
The vector potential $\bf A$ is
calculated so that
${\bf \nabla}\cdot{\bf B}=0$ at all times.
The star rotates with angular
velocity ${\bf \Omega}_*=\Omega_* ~ \hat{\bf z}$.
The intrinsic magnetic field of
the star is taken to be an aligned dipole,
with vector potential ${\bf A}={\rvecmu} \times{\bf
R}/{R^3}$.

We measure length in units of the
Bondi radius $R_B \equiv
{GM}/c_{s}^2$, with $c_s$ the
sound speed at infinity, density in
units of the density of the interstellar medium $\rho$, and
magnetic field strength in units of $B_0$ which is the field at
the pole of the numerical star.
 The magnetic moment is
measured in units of $\mu_0
=B_0 R_B^3/2$.

After reduction to  dimensionless form,
the MHD equations involve dimensionless parameters:
$$
\beta \equiv \frac{8\pi P_{0}}{B_0^2}~, \quad\quad
\tilde{\eta}_m \equiv {\eta_m \over R_B v_0} = {1\over Re_m}{~,}
\eqno(15)
$$
where  $\tilde{\eta}_m$ is the dimensionless
magnetic diffusivity,
the $Re_m$ is the magnetic Reynolds number,
and $\beta$ is
so-called gravimagnetic parameter, which describes the ratio
between pressure of the ISM medium  and magnetic field at the
poles of the ``numerical" star,  $B_0$.

Simulations were done in a cylindrical region $(Z_{min}\le z\le
Z_{max}, 0\le r\le R_{max})$. The numerical star was represented  by
a small cylindrical box with dimensions $R_*<<R_{max}$ and
$|Z_*|<<Z_{max}$. A uniform $(r,z)$ grid with $N_R\times N_Z$ cells
was used.

Initially, the density $\rho(r,z)$ and the velocity ${\bf v}(r,z)$
are taken to be constant in the region $\rho(r,z)=\rho$ and
$v=v_{*}$, $v_\phi =0$. Also, initially the vector potential ${\bf
A}$ was taken to that of a dipole so that $B_\phi=0$. The vector
potential was fixed inside the numerical star and at its surface
during the simulations. The star was initialized to be rotating at
the rate $\Omega_*$.

The outer boundaries of the computational region were treated as
follows. Supersonic inflow with Mach number ${\cal M}$ was specified
at the upstream boundary $(z=Z_{min}, 0\le r \le R_{max})$. The
variables $(\rho,~v_r,~v_z,~\varepsilon)$ are fixed. The inflowing
matter is unmagnetized with ${\bf B}=0$. At the downstream boundary
$(z=Z_{max}, 0\le r \le R_{max})$ and at the cylindrical boundary
$(Z_{min}\le z\le Z_{max}, r=R_{max}$ a free boundary condition was
applied, $\partial/\partial{\bf n}=0$. Boundary conditions are
described in greater detail in T01 and R03.

The size of the computational region for most of the simulations was
$R_{max}=0.9 R_{B}=0.9$, $Z_{min}=-R_{max}=-0.9$, with
$Z_{max}=2R_{max}=1.8$.  The grid $N_R\times N_Z$ was $289\times
865$ in most of cases. The radius of the numerical star was
$R_*=0.025{}R_{B}=0.025$ in all cases.

\section{Results}

We investigated numerically interaction of the rotating magnetar
with the ISM and the rate of spinning-down.  Simulations were done
at a variety of parameters: magnetic moments of the star $\mu$,
angular velocities $\Omega_*$, Mach numbers ${\cal M}$ and
diffusivities $\tilde{\eta}$. In most cases we varied one
parameter in a time, and kept other parameters fixed and
corresponding to the ``main" case. In the main case a star rotates
with an angular velocity $\omega_*=\Omega_*/\Omega_{K*}=0.7$,
where $\Omega_{K*}=\sqrt{GM/R_*^3}$  is Keplerian angular velocity
at the surface of the numerical star. We suggest that numerical
star (inner boundary) is much larger than the true radius of the
neutron star. Mach number is ${\cal M}=3$. We take gravimagnetic
parameter $\beta=10^{-6}$ which corresponds to the dimensionless
magnetic moment $\mu=10^{-7.5}$, and take magnetic diffusivity
$\tilde{\eta}_m=10^{-5}$.

\subsection{Dependence of Torque on  Mach Number}

For parameters corresponding
to our ``main case,'' we  did simulation
runs for the Mach numbers  ${\cal M}=1,~ 3,~ 5,~ 6,$ and  $10,$
with the ambient sound speed fixed.
   From these runs
we find that the torque decreases
with the Mach number approximately as  $\dot{L} \propto
-{\cal M}^{-0.4}$.

Figure 1a shows that for a Mach number ${\cal M}=1$
the flow is  similar to that observed in case of a
{\it non-moving} propeller (Romanova et al. 2003).
    Namely, the rapidly
rotating magnetosphere pushes matter
and magnetic flux outward in the
equatorial plane
forming the low-density, rotating torus in the
equatorial plane.
   The gravitational radius is several times larger than
magnetospheric radius so that a significant
part of the inflowing
matter is gravitationally trapped
and accumulates around the star.
   This is  similar to the case of  spherical Bondi
accretion to a star in the propeller stage.
   For ${\cal M}=1$, there is
an axial flow of matter downstream from the shock wave.
    However, the
energy-density of the inflowing matter
$\rho {\bf v}^2/2$ is smaller
than the energy-density of the equatorial
propeller outflow $\rho
{\bf v}^2/2 + {\bf B}^2/{8\pi}$ and
this is why the equatorial
structure forms.

Figure 1b shows the  flow at a larger
Mach number ${\cal M}=3$.
    In this case the energy-density
of the ISM matter is
larger than the energy-density
of the equatorial propeller-generated
wind, so that the disk structure is
bent and pushed to the direction
of the tail.
    This interaction is similar to that observed in the
simulations of magnetized
supersonic stars in the
non-rotating case (Toropina et al. 2001).
    Namely, the magnetosphere
of the star acts as an obstacle for the ISM
matter so that a bow
shock stands in front of the star and a conical shock wave forms
behind it.
     The stand-off distance and the cross-section of the
interaction is larger in the propeller case than in the
non-rotating case.
    This is because  the rotating equatorial disk of
matter and magnetic field generated by the fast
rotating magnetosphere.
     For Mach numbers ${\cal M} > 2-3$, this disk structure
is pushed by the inflowing matter into a magnetotail behind the
star.  The maximum rotational velocity of the magnetosphere
$v_{prop}=r \Omega_*$ is several times larger than the velocity of
the ISM matter, $v$.
 In common  terminology the
propeller is ``supersonic".
  In spite of this, the ISM matter pushes
this fast rotating matter together with the  magnetic field
into the magnetotail.
    Because of
angular momentum conservation, the matter in
the magnetotail continues to rotate.
   In addition, the magnetic field is twisted by the
rotating matter.  The amount of the twist depends on the parameters.
For the parameters of our main case the twist is $B_\phi/B_p\sim
0.1$.

For even larger Mach numbers, ${\cal M}=6 - 10$,  the flow is
similar to that observed for ${\cal M}=3$.
   However, the stand-off
distance is even smaller, because
the larger portion of the rotating
magnetosphere is pushed into
the magnetotail behind the star.

  The rotating magnetosphere interacts
with the non-rotating matter of
the ISM and this leads to the spinning-down of the star.
Angular momentum lost by the star flows out from the star.
    Thus, we calculate
the torque on the star by integrating the
angular momentum flux density over
a surface surrounding the star,
$$
\dot{L} = -\int d{\bf S}\cdot
\left(\rho{\bf v}_pr v_\phi -{ {\bf
B}_p r B_\phi \over 4\pi }\right)~.
$$
  Here, $d{\bf S}$ is the outward pointing surface area
element and the $p-$subscript indicates the poloidal component.
   We use a  cylindrical surface around the star
approximately at  the Alfv\'en surface to evaluate
this integral.
    Figure 2 (solid line) shows
temporal variation of this flux.
    One can see that after few initial
rotations of the star, $\dot{L}$
  becomes approximately constant.
   For comparison, we also
calculated angular momentum flux through the magnetotail at the
distance $z=0.6$ from the star. We obtained similar value of the
angular momentum flux (see dashed line at the Figure 2) because the
angular momentum  lost by the star flows into the magnetotail.

Figure 3 shows distribution of angular momentum flux densities
carried by the magnetic field $-{ {\bf B}_p r B_\phi / 4\pi }$ (top
panel) and that carried by the matter $\rho{\bf v}_pr v_\phi$
(bottom panel).
  One can see that inside the star's
magnetosphere the magnetic field gives the main contribution to the
angular momentum flux (see top panel).
  After passing the shock wave,
magnetosphere interacts with non-rotating ISM and transports part of
its angular momentum to the matter. This rotating matter propagating
to the magnetotail gives the main contribution to the  angular
momentum flux in the tail (see bottom panel).
  The rate of
angular momentum loss from the star depends on efficiency of mixing
of the ISM matter with the magnetic field of the magnetosphere.

\subsection{Dependence of the Torque on Other Parameters}

We performed a number of simulations
with different angular
velocities of the star $\Omega_*$
and different magnetic moments $\mu$.
   We calculated the total angular momentum loss rate from
the star and investigated its dependence on the $\Omega_*$ and
$\mu$.

Figure 4b shows the total angular momentum loss rate from the star
as a function of $\Omega_*$. We find that for the main case (Mach
number ${\cal M} = 3$) this dependence is $\dot{L} \propto -
\Omega_*^{1.5}$. This dependence is somewhat stronger than that
found for a non-moving star where $\dot{L} \propto - \Omega_*^{1.3}$
(Romanova et al. 2003). We expected that at Mach number ${\cal M}=1$
the torque  will have a power between 1.5 and 1.3, however, we
obtained $q=1.1$ value (see Figure 4b, rhombs). This may be
connected with different physics of interaction at very large and
very small Mach numbers.

Figure 4c shows the total angular momentum loss rate from the star
as a function of $\mu$.   Approximately, $\dot{L} \propto -
\mu^{0.6}$. This dependence is similar to that for the case of
non-moving star $\dot{L} \propto - \mu^{0.8}$ (Romanova et al.
2003).

We also calculated the dependence of the torque on diffusivity, and
obtained:  $\dot{L} \propto - \eta^{0.3}$. Diffusivity is
important factor which helps to mix non-rotating matter of the ISM
with rotating magnetosphere. It is one of important factors which
determines the amount of matter participating in spinning-down.

 In addition we calculated the dependence of the torque on the
 density of the ISM medium and obtained dependence:
$\dot{L} \propto -\rho^{0.8}$.

\subsection{Summary of Scaling Laws}

Taking into account all above dependencies, we obtain the summary of
scaling laws:

$$
{dL \over dt} \propto - ~   \eta^{0.3} \mu^{0.6} \rho^{0.8}
 {\cal M}^{-0.4} \Omega_*^{1.5}~.
\eqno(16)
$$

The dependencies do not exactly coincide with those derived in the
simple physical model (\S2.3, eq. 11), however the main tendencies
do coincide. It is important to note that in both cases the
dependence on Mach number (velocity of the star, v, in the physical
model) is {\it negative}. This is a robust result which was
understood analytically and confirmed in simulations.

We also can compare these empirical dependencies with scaling laws
proposed in \S 2.1. The dependence of $\dot{L}$ on the magnetic
moment $\mu$ implies that $\alpha \approx 1.4$.
     The dependence of $\dot{L}$ on the magnetic
diffusivity $\eta$ implies that
$\gamma \approx-0.2$.
     The dependence of $\dot{L}$ on velocity of
the star   $v$ or Mach number ${\cal M}$ implies that $\beta
+\epsilon \approx -0.8$.
     The dependence of $\dot{L}$ on the star's
rotation rate $\Omega_*$ implies that $\beta  \approx -2.4$.
     Thus equation (10) implies that
$$
{dL \over dt} \propto - ~\rho^{0.7} \mu^{0.6} \eta^{0.3}
c_s^{-3\epsilon} v^{-0.4} \Omega_*^{1.5}~. \eqno(16)
$$
Note, that dependence  on density $\rho^{0.7}$ was derived from the
scaling analysis and it is very close to that obtained from the
numerical simulations.

    The dependences we find of $\dot{L}$
on $\mu$, $v$, and $\Omega$ are not compatible with the equation (4)
of Mori and Ruderman (2003) because different approaches were used.

\section{Spinning Down of Magnetars}

Let us estimate spinning down of magnetars in real, dimensional
units. We take as a base our main dimensionless parameters: angular
velocity $\Omega / \Omega_{K*}=0.7$, Mach number ${\cal M}=3$,
gravimagnetic parameter $\beta=10^{-6}$ and magnetic diffusivity
$\tilde{\eta}_m=10^{-5}$.

Let us consider a neutron star with mass $M=1.4~M_{\odot}=2.8
\times 10^{33}$ g and radius $R_{NS}=10^6$ cm. The density of the
ambient interstellar matter is taken to be $\rho=1.7 \times
10^{-24}$ ~g$/$cm$^3$ ($n=1/$~cm$^{3}$). The sound speed in the
ISM, $c_s=30$ km/s and a star moves in the interstellar medium
with velocity $v=3 c_s=90$ km/s.

Using the definitions $\beta=8\pi p/B_0^2$, $p=\rho c_s^2/\gamma$,
we obtain the magnetic field at the surface of the ``numerical" star
$B_0 = (8 \pi /{\gamma}{\beta})^{1/2} \rho ^{1/2} c_s  \approx
0.15~{}n_{1} c_{30}$ G.

We use the Bondi radius as a length scale. This radius is $$ R_B
=GM_*/c_s^2 \approx 2.1\times 10^{13}c_{30}^{-2} ~{\rm cm}~. $$

Recall that the size of the numerical star is $R_*= 0.025
R_B\approx 5.2 \times 10^{11}c_{30}^{-2}{\rm cm}$ in all
simulations. Thus the magnetic moment of the numerical star is
$\mu =B_0 r_*^3/2\approx 1 \times 10^{33}n_{1} c_{30}^{-5}$
Gcm$^3$. Note that this is of the order of the magnetic moment of
a real neutron star with surface field $B_{NS}=10^{15}$ G and
radius $R_{NS}=10^6$ cm. So we can suggest that a real neutron
star is hidden inside numerical star. The neutron star with so
strong magnetic field is considered as a magnetar.

The numerical star in our main case rotates with angular velocity
$\Omega_*=0.7 \Omega_{K*}=2.5 \times 10^{-5}~{ s^{-1}}$ and this
corresponds to a period of $P_*=2\pi/\Omega_* \approx 2.5\times
10^5$ s.

For these parameters the corotation radius, $R_{cor}=
(GM/{\Omega_*}^2)^{1/3} = R_*/{\omega_*}^{2/3} \approx 1.3 R_*$,
is appreciably less than $R_{m} \approx 5R_*$. The light cylinder
radius $R_L=cP/2 \pi = 1.17 \times 10^{15}P_5~{\rm cm}$ is greater
than the radius of magnetosphere taken from simulations $R_{m}
\approx 2.5 \times 10^{12}c_{30}^{-2}{\rm cm}$. So this case
$R_{cor} \le R_{m} \le R_L$ corresponds to the propeller regime.

Therefore our simulations model an isolated magnetar which spins
down by the propeller effect.

Now we can estimate a time of spin down due to the propeller
torque using equation (9). An angular momentum of a star is equal
to $L_*=I_* \Omega_*$, where $I_*=2/5 M_* R_*^2$ - moment of
inertia for sphere. Taking $I_* \approx 10^{45}~{\rm g cm^2}$ and
$\Omega_*=2 \pi /P_*$, we obtain $L_*=2 \pi I_* / P_* \approx 3
\times 10^{40} P_5^{-1}~{\rm g cm^2/s}$.

The angular momentum loss rate $\dot{L}=\tilde{\dot{L}}
\dot{L_0}$, where for typical simulation run we find
$\tilde{\dot{L}} \approx 10^{-7}$. Taking into account dimensional
parameter $\dot{L_0} \approx -3\times 10^{36} n_{1}
c_{30}^{-4}~{\rm g (cm/s)}^2$ we obtain $\dot{L} \approx -3\times
10^{29} n_{1} c_{30}^{-4}~{\rm g (cm/s)}^2$.

In reality $\dot{L}$ is not constant and depends on many
parameters - velocity of the star, angular velocity and magnetic
field of the neutron star. We performed a number of simulations
with different angular velocities $\Omega_*$ and different
magnetic moments $\mu$ of the star and find dependencies: $\dot{L}
\propto \Omega_*^3 \mu^{0.6}{\cal M}^{-0.4}$, or $\dot{L} \propto
P_*^{-1.5} \mu^{0.6}{\cal M}^{-0.4}$. Thus,
$$
\dot{L} \approx 3
\times 10^{29} n_{1} c_{30}^{-4}B_{15}^{0.6}P_5^{-1.5}{\cal
M}_3^{-0.4}~{\rm g (cm/s)}^2~.
\eqno(17a)
$$
Thus the characteristic  spin-down time is
$$
T = L_*/\vert
\dot{L} \vert \approx 10^4~n_{1}^{-1} c_{30}^{4}
B_{15}^{-0.6}P_5^{0.5}{\cal M}_3^{0.4}~{\rm yr}~.
\eqno(17b)
$$
Now we can estimate the time-scale of evolution at the propeller
stage. For periods $P_*=10^3$ s, which correspond to beginning of
the propeller stage, the evolution scale will be $\Delta T =10^3$
years, while at period $P_*=10^6$ s corresponding to the end of
propeller stage $ T =3 \times 10^4$ years. Thus we see that
magnetars are expected to spin down very fast at the propeller
stage. This time-scale however may be much larger if diffusivity is
very small.

\subsection{Restrictions of the Model}

The axisymmetric resistive MHD numerical code used in this paper is
a robust and has been checked at a number of tests and also it has
been used for solution of similar, but somewhat simpler problems,
such as Bondi accretion to a non-moving star, wind accretion to a
moving magnetized non-rotating star, and others. This paper is a
logical continuation of the series of simulations with increased
complexity. The bow shock which forms in our simulations is similar
to one obtained in simulations of the Earth's magnetosphere
interacting with the solar wind (e.g., Gombosi, Powel, \& van Leer
2000) and simulations of the pulsar wind interacting with
magnetosphere of the companion neutron star (Arons et al. 2005).

However, there is one factor which is important and basically is not
known, this is the diffusivity. The role of diffusivity was
underestimated in the analytical papers on propellers, however it is
very important for mixing of the incoming  matter to the magnetic
field of the  magnetosphere. In our model we took some value of
diffusivity such that part of the incoming matter diffuses through
the field lines and takes part in the spinning-down of the star.
However, if the diffusivity will be much smaller, then the
time-scale presented by the formulae (17b) will increase.

\section{Discussion of Possible Magnetar candidates}

It is not clear whether we can observe old, slowly rotating
magnetars. One of unknown aspects is that we do not know whether
magnetic energy will continue to release, and at  which rate, and
what is the expected spectrum of such radiation. Theory of radiation
from relatively fast rotating magnetars has been recently developed
(e.g., Beloborodov \& Thompson 2006), which predicts radiation from
the whole magnetosphere mostly in the X-ray.  However we do not know
what would be the luminosity and spectrum of much slower rotating
magnetars with periods  $P < 10^3$ seconds and thus it is not clear
how to find them, even if many of them survive.  This is an obstacle
in searching the old magnetars even with the best modern
instruments. There are, however, candidates for magnetars: SGRs and
AXPs and also isolated neutron star (INS) candidate RX J1856.5-3754.

\subsection{ Clustering of Periods of SGRs and AXPs} One of amazing
properties of SGRs and AXPs, which are strong candidates for
magnetars, is the clustering of periods around 6-12 seconds. This is
still a stage when the light cylinder is smaller than magnetospheric
radius, and the processes at the light cylinder may be responsible
for the spin-down.   The estimated age of these objects is around
$10^3-10^4$ years which is an evidence of their fast spinning down
at this stage.  Analysis of possible evolution of periods have shown
that after $P>10$ seconds, these objects should either spin down
much faster, or the magnetic field should decay very fast (Psaltis
\& Miller 2002, see also Colpi et al. 2000). The spinning-down at
this stage has not been investigated yet. Recent relativistic 3D
numerical simulations of aligned and misaligned rotators and further
modeling in this direction  may help to understand what is the
expected spinning-down rate at the pulsar-type stages (Spitkovsky
2004; 2006). Unfortunately, we can not extrapolate formulae (17b) to
much shorter periods, $P\sim 10$ seconds, because our results are
only relevant to propeller stage with $P > 10^3$ seconds. Our
simulations show, that propeller mechanism is very efficient in
spinning-down of magnetars. However, we should remember, that there
is a big uncertainty in the results of simulations: we do not know,
which part of the ISM matter from the cross-section $\sigma = \pi
r_{st}^2$ will participate in the spinning-down of the star. This
greatly depends on the diffusivity, which is not known. In spite of
that it is important to know the dependencies of the torque on
different parameters. So, our model can not explain the clustering,
however if a magnetars with period $P > 10^3$ seconds will be
discovered in the future, then these formulae will describe the
subsequent evolution.

\subsection{\bf Isolated Neutron Star candidate RX J1856.5-3754.}

Of the small number  of  isolated neutron star candidates, the
object RX J1856.5-3754 is a special  because it has a
$H_{\alpha}$ nebulae with the shape of a shock wave (van Kerkwijk \&
Kulkarni 2001).
  The origin of this nebulae as well as the origin of
the INS candidate are not yet known.
  One  possibility is that it
is a misaligned pulsar with magnetic field $B\sim 10^{12} -
10^{13}~{\rm G}$  and period $P\sim 1 {\rm s}$.
 Then the nebulae may
be a shock wave which appears as a result of the interaction of the
relativistic wind of the pulsar with the ISM (e.g., van Kerkwijk and
Kulkarni 2001; Romanova et al. 2001; Braje \& Romani 2002;
Romanova, Chulsky, \& Lovelace 2005).
  A  weak part of this model is the lack of pulsations.
 The X-ray pulsed fraction is $< 1.3\%$ (Burwitz et al. 2003).
  Another idea is that this is an old neutron star which had
a recent accretion event and is slowly cooling down. Then the
$H_{\alpha}$ nebulae may be connected with the ionization by the ISM
(van Kerwijk \& Kulkarni 2001).

Another possible idea is that RX J1856.5-3754 is a relatively old,
slowly rotating magnetar (Romanova et al. 2001) which
spun-down rapidly during its the propeller stage (Mori and Ruderman
2003).
 This model may explain the lack of pulsations.
  However, there is no
simple explanation of the $H_{\alpha}$ nebulae.
   In spite of
the large cross-section of the bow-shock, the energy of the ISM
released at the shock $ \dot E_{shock}\approx 0.5\rho v^3\pi
r_m^2\approx 5\times 10^{23} n^{2/3} v_7^{7/3} B_{15}^{2/3}~{\rm
ergs~s^{-1}} $ is much smaller than that of the $H_{\alpha}$ nebulae.
   Also, reconnection in the magnetotail
(Toropina et al. 2001)  gives insufficient power.
An additional
heating mechanism of the star or magnetosphere is necessary to
explain the $H_{\alpha}$ nebulae.
 Heating of the star due to
accretion from the ISM  is
excluded because  very little
matter  accretes to a strongly magnetized star
owing  to the very narrow
accretion columns (Toropina et al. 2003).
   There is the possibility of
heating of the star by the release of the magnetic energy (e.g.,
Beloborodov \& Thompson 2006).
  However, the magnetic heating of
slowly-rotating magnetars has not been investigated.
Thus the powering the $H_\alpha$ nebulae is uncertain but
not ruled out.

\section{Conclusions}

  Using axisymmetric MHD simulations
we have studied the supersonic propagation through the ISM of
magnetars in the propeller stage.
     We have done many simulation runs for the
purpose of determining the angular momentum
loss rate of the star due to the interaction
of its magnetosphere with the shocked ISM.
    We conclude, that the interaction may be highly effective
in spinning-down  magnetars.
   A star with
magnetic field $B \sim 10^{13}-10^{15} G$ is expected to spin-down
in $\Delta T \sim 10^4-10^5 {\rm years}$.
  This time may be longer if
the ISM material does not efficiently interact with the external
regions of the magnetar's magnetosphere.
    Therefore, after relatively short stages of
pulsar and propeller activity, a magnetar becomes a very slowly
rotating object, with  a period $P > 10^5-10^6 s$, which is much
longer than the periods expected for ordinary pulsars.
   This may be a
reason why the number of soft gamma repeaters, which are candidate
 magnetars, is so small.
   We should note however, that the rate of
spinning-down depends on the  magnetic diffusivity
which is not known.
    At lower diffusivity the rate of
spinning-down will be lower. The
INS candidate RX J1856.5-3754 may be an example of a slowly rotating
magnetar.
 However,  this model does not explain the
$H_{\alpha}$ nebulae.
  An ordinary misaligned pulsar explains the
different features more easily,  excluding the fact that no
periodic fluctuations were observed from this object.

\section*{Acknowledgements}
This work was supported in part by NASA grants NAG5-13220,
NAG5-13060, by NSF grant AST-0507760 and by Russian program
``Astronomy''. We thank anonymous  referee for valuable suggestions,
Dr. V.V. Savelyev for the development of the original version of the
MHD code used in this work and Dr. Yuriy Toropin for discussions.


\begin{thebibliography}{10}


\bibitem{aro04} Arons, J., Backer, D.C., Spitkovsky, A., \&
Kaspi, V.M. 2005, ASP Conference Series, vol. 328, p. 95, Eds. F.A.
Rasio and I.H. Stairs (astro-ph/0404159)


\bibitem{} Beloborodov, A.M., \& Thompson, C. 2006, ApJ, in press
(astro-ph/0602417)

\bibitem{} Braje, T.M., \& Romani, R.W. 2002, ApJ, 580, 1043

\bibitem{} Burwitz, V., Haberl, F., Neuh{\"a}user, Predehl, P.,
Tr{\"u}mper, J., \& Zavlin, V.E. 2003, A\&A, 399, 1109

\bibitem{col00} Colpi, M., Geppert, U., \& Page, D. 2000,
ApJ, 529, L29

\bibitem{dav73} Davidson, K., \& Ostriker, J.P. 1973, ApJ, 179, 585

\bibitem{} Davies, R.E., Fabian, A.C., \& Pringle, J.E. 1979, MNRAS, 186,
779

\bibitem{} Davies, R.E., \& Pringle, J.E. 1981, MNRAS, 196, 209


\bibitem{dun92} Duncan, R.C., \& Thompson, C. 1992, 392, L9


\bibitem{} Gombosi, T.I., Powell, K.G., \& van Leer, B. 1999, JGR, 105,
Issue A6, 13141

\bibitem{hard92} Harding, A.K. \& Leventhal, M. 1992, Nature, 357, 388

\bibitem{hurl99} Hurley, K., et al. 1999, Nature 397, 41

\bibitem{ikh} Ikhsanov, N.R. 2002, A\&A, 381, L61

\bibitem{ill75} Illarionov, A.F., \& Sunyaev, R.A. 1975. A\&A, 39, 185

\bibitem{kou94} Kouveliotou, et al. 1994, Nature, 368, 125

\bibitem{kou98} Kouveliotou, et al. 1998, Nature, 393, 235

\bibitem{kou99} Kouveliotou, et al. 1999, ApJ, 510, L115

\bibitem{kul93} Kulkarni, S.R. \& Frail, D.A. 1993, Nature, 365, 33

\bibitem{lip92} Lipunov, V.M. 1992, {\it Astrophysics of Neutron
Stars}, (Berlin: Springer Verlag)

\bibitem{lov99} Lovelace, R.V.E., Romanova, M.M., \& Bisnovatyi-Kogan,
G.S. 1999, ApJ, 514, 368

\bibitem{mori03} Mori K., Ruderman M.A. 2003, ApJ, 592, L75


\bibitem{} P\'erez-Azor\'in, J.F., Miralles, J.A., \& Pons, J.A.
2005, A\&A, 433, 275

\bibitem{} Psaltis, D. \& Miller, M. C. 2002, ApJ, 578, 325



\bibitem{rap04} Rappaport, S. A., Fregeau, J. M., \& Spruit, H.
2004, ApJ, 606, 436

\bibitem{rom01} Romanova, M.M., Toropina, O.D., Toropin, Yu.M.,  \&
Lovelace, R.V.E. 2001, AIP conference proceedings, Vol. 586. Eds.
J.C. Wheeler and H. Martel, p.519

\bibitem{rom03} Romanova, M.M., Toropina, O.D., Toropin, Yu.M.,  \&
Lovelace, R.V.E. 2003, ApJ, 588, 400

\bibitem{rom04} Romanova, M.M., Ustyugova, G.V., Koldoba, A.V., \&
Lovelace, R.V.E. 2004, ApJ, 616, L151

\bibitem{rom05} Romanova, M.M., Ustyugova, G.V., Koldoba, A.V., \&
Lovelace, R.V.E. 2005, ApJ, 635, L165

\bibitem{} Romanova, M.M., Chulsky, G.A., \& Lovelace, R.V.E. 2005,
ApJ, 630, 1020

\bibitem{rut01} Rutledge, R.E. 2001, ApJ, 553, 796

\bibitem{shv70} Shvartsman, V.F. 1970, Radiofizika, 13, 1852

\bibitem{} Spitkovsky, A. 2004, IAU Symp. 218, 357, astro-ph/0310731

\bibitem{spi06} Spitkovsky, A. 2006, ApJ Letters (astro-ph/0603147)

\bibitem{tho95} Thompson, C., \& Duncan, R.C. 1995, MNRAS, 275, 255

\bibitem{T99} Toropin, Yu.M., Toropina, O.D., Savelyev, V.V.,
Romanova, M.M., Chechetkin, V.M., \& Lovelace, R.V.E. 1999, ApJ, 517, 906

\bibitem{T01} Toropina, O.D., Romanova, M.M., Toropin, Yu.M., \&
Lovelace, R.V.E. 2001, ApJ, 561, 964

\bibitem{T03} Toropina, O.D., Romanova, M.M., Toropin, Yu.M.,
\& Lovelace, R.V.E. 2003, ApJ, 593, 472

\bibitem{} Ustyugova, G.V., Koldoba, A.V., Romanova, M.M., \&
Lovelace, R.V.E. 2006, ApJ, accepted (astro-ph/0603249)


\bibitem{} van Kerkwijk, M.H., \& Kulkarni, S.R. 2001, A\&A, 380,
221

\bibitem{vas97} Vasist, G., \& Gotthelf, E. V. 1997, ApJ, 486, L129

\bibitem{} Wang, Y.-M., \& Robertson, J.A. 1985, A\&A 151, 361

\bibitem{zhu93} Zhukov, V.T., Zabrodin, A.V., \&
Feodoritova,~O.B. 1993, Comp. Maths. Math. Phys., 33, No. 8, 1099


\end{thebibliography}
\end{document}